\documentclass[final]{svjour3}
\usepackage[dvipdfmx]{graphicx}
\usepackage{rotating}
\usepackage{amssymb}
\usepackage{mathptmx}
\usepackage{amsmath}
\usepackage[numbers]{natbib}
\usepackage{xcolor}
\usepackage{hyperref}
\hypersetup{
    colorlinks=true,
    linkcolor=blue,
    filecolor=magenta,      
    urlcolor=cyan,
    pdftitle={Overleaf Example},
    pdfpagemode=FullScreen,
    }
\usepackage[normalem]{ulem}

\makeatletter
\journalname{Journal of Low Temperature Physics}

\bibpunct{}{}{,}{s}{}{,}

\begin{document}

\newcommand{\hdblarrow}{H\makebox[0.9ex][l]{$\downdownarrows$}-}
\title{Optical leakage mitigation in ortho-mode transducer detectors for microwave applications}

\author{$^1$R.~Gualtieri \and $^1$P.~S.~Barry \and $^1$T.~Cecil \and $^{1,2}$A.N.~Bender \and $^{1,2}$C.L.~Chang \and $^{2,3}$J.C.~Hood \and $^1$M.~Lisovenko \and $^1$V.G.~Yefremenko}
\institute{$^1$HEP, Argonne National Laboratory, Lemont, IL 60439, USA\\ 
           $^2$KICP, Department of Astronomy and Astrophysics, University of Chicago, Eckhardt Research Center 5640 South Ellis Avenue Chicago, IL, 60637\\
           $^3$Department of Astronomy and Astronomy, Vanderbilt University, 6301 Stevenson Center, Nashville, Tennessee 37240}


\maketitle
\begin{abstract}
Planar ortho-mode transducers (OMTs) are a commonly used method of coupling optical signals between waveguides and on-chip circuitry and detectors. While the ideal OMT-waveguide coupling requires minimal disturbance to the waveguide, when used for mm-wave applications the waveguide is typically constructed from two sections to allow the OMT probes to be inserted into the waveguide. This break in the waveguide is a source of signal leakage and can lead to loss of performance and increased experimental systematic errors. 
Here we report on the development of new OMT-to-waveguide coupling structures with the goal of reducing leakage at the detector wafer interface. 
The pixel to pixel optical leakage due to the gap between the coupling waveguide and the backshort is reduced by means of a protrusion that passes through the OMT membrane and electrically connects the two waveguide sections on either side of the wafer. 
High frequency electromagnetic simulations indicate that these protrusions are an effective method to reduce optical leakage in the gap by $\sim80\%$ percent, with a $\sim60\%$ filling factor, relative to an standard OMT coupling architecture. 
Prototype devices have been designed to characterize the performance of the new design using a relative measurement with varying filling factors. 
We outline the simulation setup and results, and present a chip layout and sample box that will be used to perform the initial measurements.
\end{abstract}

\section{\label{sec:context}Introduction}
Experiments that seek to measure the power and or polarization of microwave signals need small polarization errors and exquisite control of the instrument systematics. A popular method for coupling microwave signals to on wafer detectors utilizes feedhorns because of their well constrained beam shape and very small polarization errors as has been demonstrated in the field by a number of CMB experiments \cite{2012Bleem, 2016SPIE.9914E..0VH,2016SPIE.9914E..16S}.
Any unwanted coupling (e.g., to the wrong detector, or to the correct detector but with the wrong phase) results in a small loss of efficiency and a potentially much more serious systematic error in polarization. 
Signals received by a horn antenna are typically coupled to circular waveguide and then to a planar transmission line on the detector wafer using an ortho-mode transducer, which in made up of two pairs of orthogonal probes in a cross configuration. 
The probes must penetrate the waveguide wall, without spoiling the mode content in the waveguide. 
A thin silicon nitride membrane can be used to support the probes inside the waveguide, but this requires a discontinuity in the waveguide wall. 
Optical power can leak out of the gap and propagates either along the vacuum gap, or couples into the high-dielectric-constant detector wafer. 
Adding resonant choke structures around the waveguide gap helps reduce the leakage within the vacuum gap across a narrow bandwidth, but it is difficult to reduce the leakage below a few percent \cite{2015Duff, 2016Hubmayr}. 
In an attempt to address this issue, groups have implemented strategies that include absorber-filled trenches around the waveguide to capture stray power, but these absorbing materials are not compatible with planar wafer detector fabrication processes, so it is difficult to put the absorber where it would be most effective \cite{2016Shibo, 2010Niemack}.

We develop a different approach based on a new ortho-mode transducer structure that minimizes the effect of the gap in the waveguide wall. Electrical connection is made across the gap by means of a protrusion from the backshort that connects with the feedhorn stack, passing through a set of perforations in the silicon nitride (SiN) supporting membrane. The electrical connection is guaranteed by careful design and ensuring a gold-plated surface. 
With this structure the only discontinuities seen by the propagating modes in the waveguide are the apertures needed for the OMT SiN supporting legs. 
The width of the aperture is designed to guarantee structural stability and minimal leakage given the incoming radiation frequency.  
This approach is relatively simple and can be added to existing detector designs with only minor changes.

\section{\label{sec:EMsim}Electromagnetic Simulation}
\begin{figure}[htbp]
\begin{center}
\includegraphics[width=1.0\linewidth, keepaspectratio]{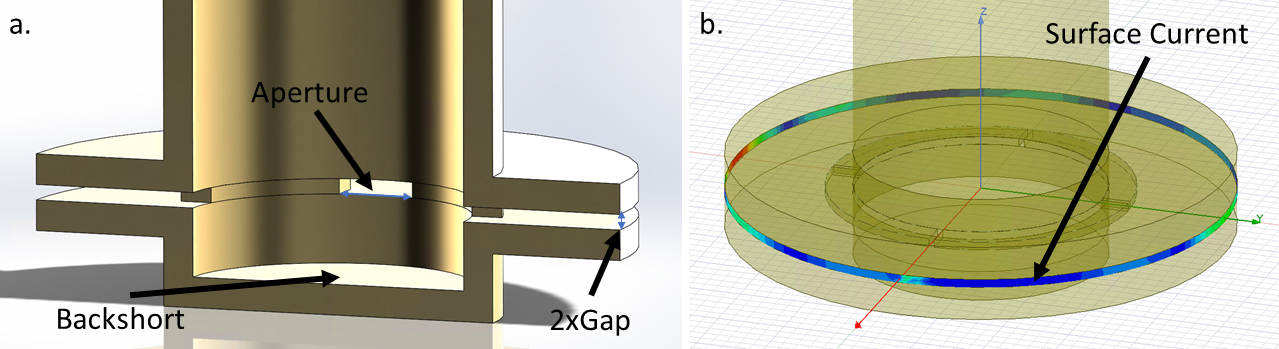}
\end{center}
\caption{Model used for electromagnetic simulations: \textbf{a.} Full model showing input waveguide, protrusions, and backshort. Several combination of  gap and aperture width parameters are explored. \textbf{b.} The HFSS model with emphasis on the dummy surface where the surface current is integrated to evaluate the leakage. }
\label{WG_cartoon_1}
\end{figure}

We first evaluate the potential performance of the waveguide protrusions through EM simulations using the Ansys HFSS \footnote{\url{https://www.ansys.com/products/electronics/ansys-hfss}} simulation package. 
The basic simulation model consists of an input waveguide, quarter wavelength backshort, and protrusions to electrically connect the waveguide and backshort as shown in Fig.~\ref{WG_cartoon_1}. 

To model the protrusion we designed a circular ring to perfectly fill the space between the horn stack and the backshort assembly with its height defined as twice the gap, $h = 2\times gap$. The waveguide is modeled as a perfect conductor with inner diameter of 0.073~in (1.8542mm), which is based upon commercially available 150~GHz feedhorns \footnote{\url{https://custommicrowave.com/Circular-Horns-With-Circular-Input}}.  
The wall thickness is set to 7.87~mils (0.200mm), although in a physical device this thickness can be increased as needed based upon fabrication needs. The backshort is at $\lambda /4$ distance from the center of the assembly, nominally where the detector wafer would sit. The protrusion connects the waveguide and backshort assemblies by means of four annular sections separated by apertures.  The apertures provide an opening in the waveguide that the legs supporting the OMT membrane and microstrip transmission lines pass through. The slots in the waveguide are designed as orthogonal cuts into a circular section defined by two radii: the internal waveguide radius,$r_1$, and the external one defined as the internal radius plus the WG wall thickness, $r_2$.
The angles from the center of the waveguide intersecting the aperture are defined as
\begin{equation}
    \theta_n = 2arcsin\Big(\frac{AW}{2r_n}\Big),
\end{equation}
where AW is the aperture width. The area of each of the four slots can be then calculated as
\begin{equation}
    A_{slot}= A_0-A_1-A_2,
\end{equation}
where 
\begin{equation}
    \begin{gathered}
    A_0=AW\Big(r_2\cos{\frac{\theta_2}{2}}-r_1\cos{\frac{\theta_1}{2}}\Big),\\
    A_1=r_1^2/2(\theta_1-\sin(\theta_1)\quad\mbox{and}\\ A_2=r_2^2/2(\theta_2-\sin(\theta_2).    
    \end{gathered}
\end{equation}

The fractional filling factor, ($ff$), as a function of the aperture width is then:
\begin{equation}\label{ff}
    ff(AW)=100\times\Big[1-\frac{4A_{slot}}{\pi r_2^2-\pi r_1^2}\Big]
\end{equation}
All the assembly features are modeled as perfect conductors with perfect electrical connection between each other.
The model evaluates two primary parameters - the gap between the upper waveguide and lower backshort that the OMT membrane occupies and the aperture. Power is injected in the waveguide as a wave-port, perpendicular to the central axis of the waveguide, with the first two modes enabled. A dummy cylindrical surface is placed outside of the assembly with surface impedance of $377~\mathrm{\Omega}$
set to match free space. The surface current density is integrated to determine the power that has leaked through the gap between the waveguide and backshort. The aperture width and the gap height are varied in the simulation and the frequency is swept between 100 and 180~GHz.  

\begin{figure}[htbp]
\begin{center}
\includegraphics[width=1.0\linewidth, keepaspectratio]{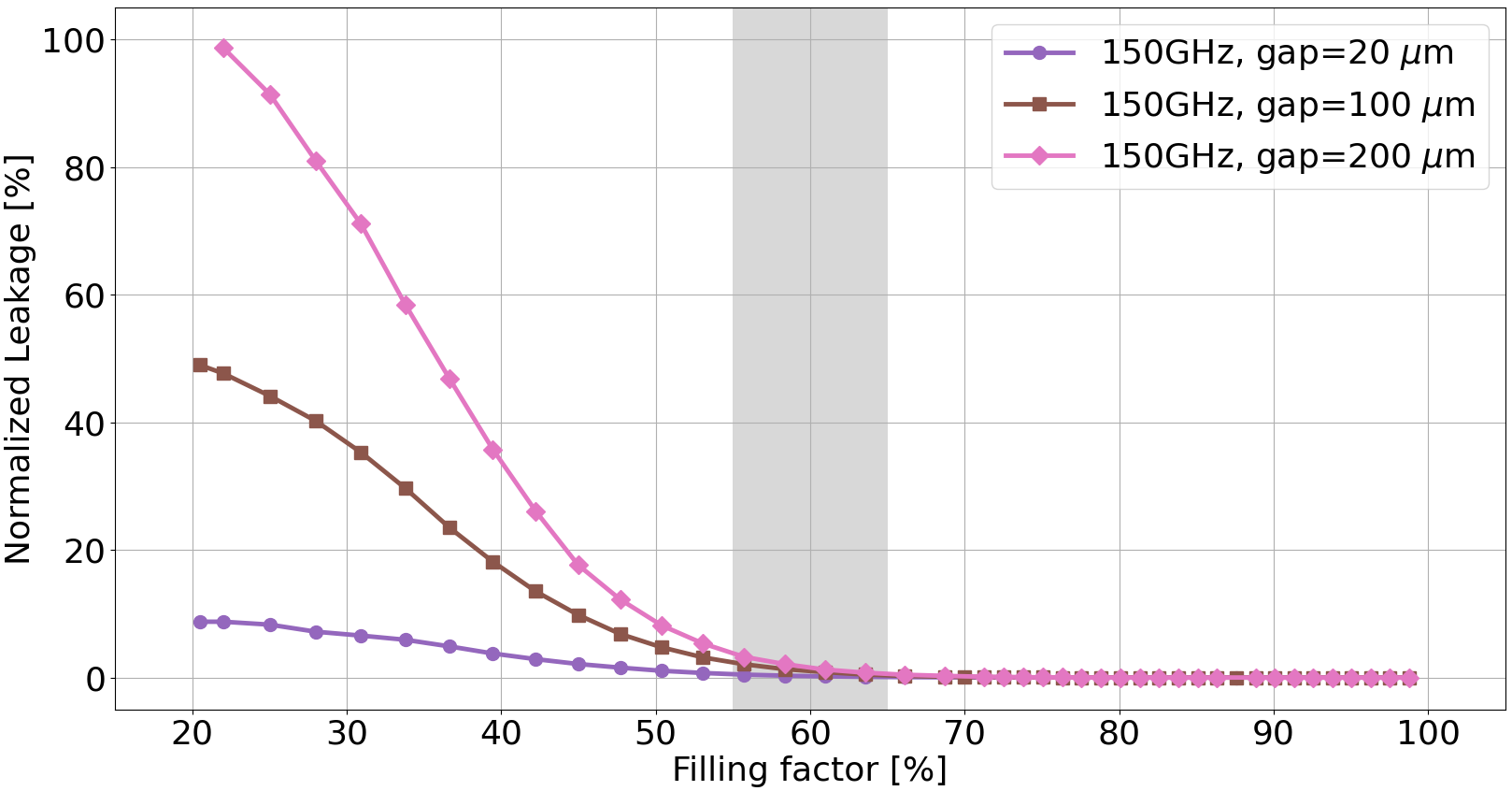}
\caption{Normalized leakage as a function of the protrusion filling factor. The results shown have being normalized to the extrapolated leakage value in the case of no protrusion and a 200~µm gap for three different gaps: 20, 100 and 200~µm. The range has been chosen to capture the behavior of future small gap designs, a fairly common and standard 100~µm design and a leaky by construction design at the chosen 150~GHz center band frequency. Discussion on the results in text.}
\end{center}
\label{results}
\end{figure}

The simulation explores the parameter space on a grid of values for the aperture width and the gap.
The parameters have been chosen to represent combinations allowed by the fabrication processes discussed below.
The gaps are swept between $20$ and 200 µm and the aperture width from 20 to 1500 µm.\\
The key results have been summarized in Fig.~\ref{results}, where the normalized leakage is shown as a function of the filling factor. The leakage has been normalized dividing the data by the value of the leakage with no protrusion (corresponding to zero filling factor) and the largest gap investigated (200 µm). From these data we extrapolate the fractional reduction of leakage.

It is clear from Fig.~\ref{results} that reducing the gap reduces the leakage.  Introducing the protrusions reduces the leakage with a rapid reduction in leakage as the filling factor increases. For a filling factor of 60$\%$ the normalized leakage is below 2$\%$ for all the gap heights shown. Thus by introducing the protrusions it is possible to reduce the overall leakage level or maintain the same level of leakage with reduced tolerances for the gap height.

\section{Experimental Design}
\begin{figure}[htbp]
\begin{center}
    \includegraphics[width=1.0\linewidth,keepaspectratio]{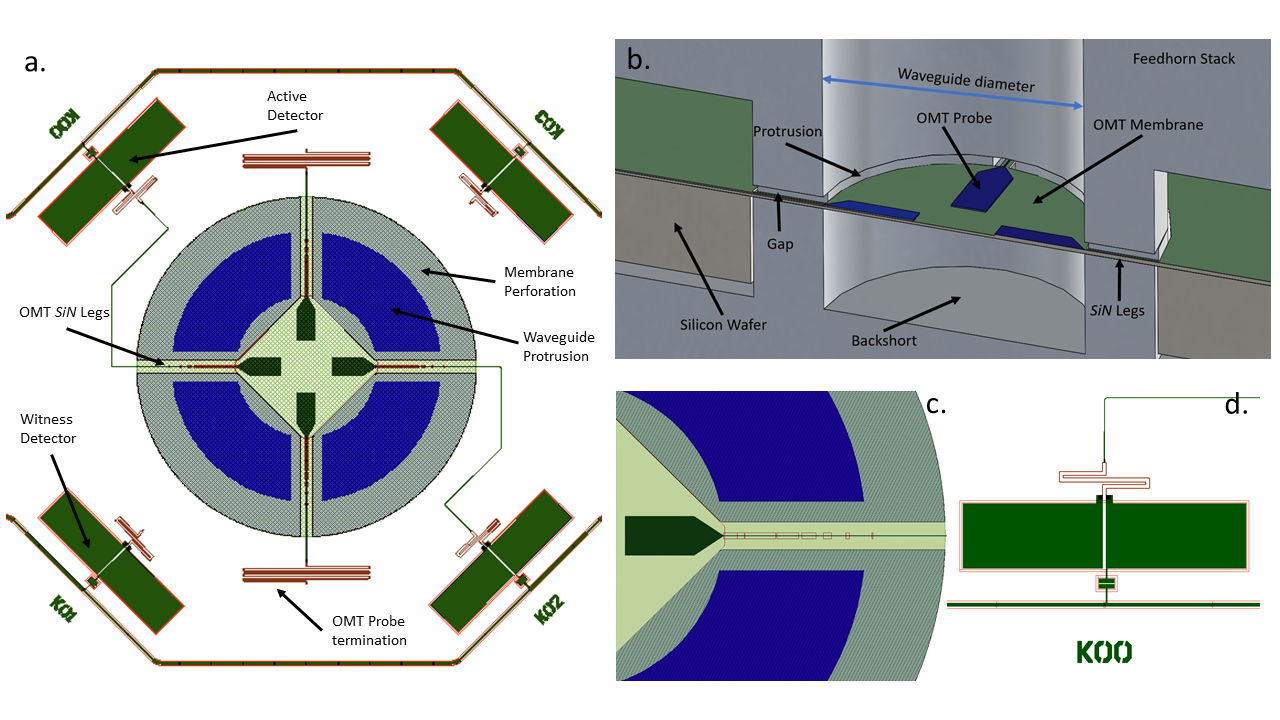}
\end{center}
\caption{\textbf{a.} Schematic showing the test chip with four OMT membranes and KIDs arranged around an in-between the OMTs. \textbf{b.} render of a CAD of a one pixel implementation we used to study the mechanical feasibility and thermal properties of the assembly. \textbf{c.} A closeup of a single OMT membrane, and the \emph{SiN} leg supporting the OMT membrane showing the clearance between the protrusions and the membrane. \textbf{d.} Detector layout closeup.}
\label{OMT_perforation_1}
\end{figure}

We have designed a test chip based upon the simulation results to experimentally characterize the impact of adding electrical protrusions through the OMT membrane. The layout of the test chip is shown in Fig.~\ref{OMT_perforation_1}. 

\begin{figure}[htbp]
\begin{center}
\includegraphics[width=1.\linewidth, keepaspectratio]{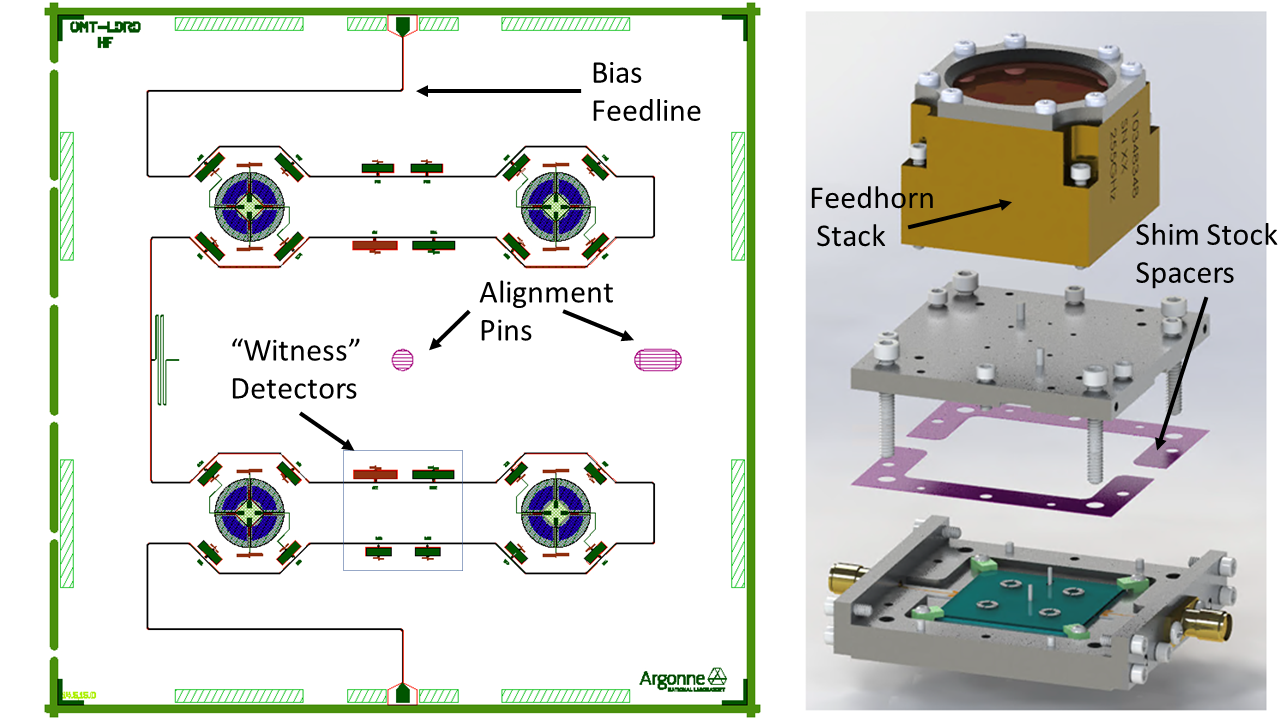}
\end{center}
\caption{Four pixel prototype layout and 3D cad model of the detector's holder, the aluminum box lid will be adapted to connect commercial feedhorns for cryogenic optical tests, high precision alignment will be provided by slots and pins. The gap will be tuned precisely using shim stock spacers.\\
\textbf{\textit{Left}}: Mask layout for the four pixel prototype wafer. \textbf{\textit{Right}}: Exploded view of the full proposed assembly.}
\label{pixel}
\end{figure}

The test chip consists of four pixels, each with an OMT membrane perforated by four annular shaped cutouts in the SiN layer. Four legs remain to support the OMT probes. The legs are 200~µm wide, which leaves sufficient space for the coplanar waveguide-to-microstrip transition and has been found to be robust during the fabrication process. With this configuration a filling factor of up to 80\% is possible. 
At a filling fraction of 75\% the aperture width is close to 350~µm, which allows for 75~µm of space between the protrusions and the membrane. 
This clearance is sufficient to allow for fabrication tolerances in the machining of the sample box, alignment tolerance between the chip and sample box, and compensation for differential thermal contraction between the silicon wafer and the Al sample box, Fig.~\ref{pixel}.
Two of the OMT probes are terminated with a pair of kinetic inductance detectors (KIDs) to measure the power coupled to the probes, while the second pair is terminated with a resistive absorber (the planned measurement setup will use a non-polarized blackbody source so only one set of probes is needed). The other two KIDs around each membrane are used as witness sensors to measure stray light directly outside the waveguide area. KIDs are known to be highly susceptible to stray light within the substrate \cite{2018Yates} and so will be used a proxy for the power leakage.
As shown in the left panel of Fig.~\ref{pixel}, in between the four membranes are eight KIDs  that serve as witness detectors for stray light between the four OMT membranes.
The test chip will be paired to a sample box as shown in Fig.~\ref{pixel} \emph{right}). The sample box consists of an aluminum enclosure with four apertures matching the waveguide diameter on the lid and four machined back-shorts on the bottom. One of the four back-shorts will be of nominal length, $\lambda/4-g$, where $g$ is the gap, this will be our regular waveguide used as a reference. The remaining three will have a longer WG section, $\lambda/4+2\cdot g$, effectively acting as the simulated protrusion. The three protruded WGs will have cuts to explore the 25\%, 50\% and, 75\% filling factors.

The box lid will be connected to a custom feedhorn stack and filter holder as shown in Fig.~\ref{pixel}, which have the same waveguide apertures used in the simulation. 
We plan several cryogenic runs illuminating the probes with a cryogenic black body source, a number of differential tests will be possible removing the feedhorns and blinding the input waveguides.

\section{Conclusions}
An efficient way to mitigate the optical leakage in close packed arrays of microwave detectors is necessary to meet future experimental needs. 
We demonstrate with simulations that adding protrusions to electrically connect the two halves of the sample box is a viable path to reduce the leakage efficiently without requiring very small gap heights or choke structures. A suppression of a factor of $\mathcal{O}(2)$ is achievable with a filling factor of $~60\%$. This filling factor can be achieved using cutouts in the OMT membrane that can be etched without compromising the structure stability of the membrane. The protrusion is relatively easy to machine, removes the need for choke structures and offers a potential path to scaling for a large number of pixels \cite{2021Abitbol}.\\ 
Future work will include fabricating devices with smaller apertures, production of the optical sample holder, and  cryogenic testing. With this solution and an optimized choice of the parameters, the radiation will see an essentially continuous waveguide. This will maximize the coupling with the probe and strongly suppress the leakage allowing a significant mitigation of the pixel to pixel leakage systematics.\\
We plan to extend the analysis to higher frequencies, above 200~GHz, as we expect the leakage through the gap to be more prominent as the radiation frequency increases and the gap becomes commensurable with the radiation wavelength. 

\section{Acknowledgments}
Work at Argonne, including use of the Center for Nanoscale Materials, an Office of Science user facility, was supported by the U.S. Department of Energy, Office of Science, Office of Basic Energy Sciences and Office of High Energy Physics, under Contract No. DE-AC02-06CH11357.
Ansys: \url{https://www.ansys.com/products/electronics/ansys-hfss}




\begin{thebibliography}{99}
\bibitem{abazajian2019cmbs4}
    CMB-S4 Science Case, Reference Design, and Project Plan, K.~Abazajian et al., 2019, 1907.04473
\bibitem{2010SPIE.7741E..22H}
    Optical Efficiency of Feedhorn-Coupled TES Polarimeters for
Next-Generation CMB Instruments, J.W.~Henning et al., 2010, doi: 10.1117/12.859478
\bibitem{2012Bleem}
    An Overview of the SPTpol Experiment, L.~Bleem et al., 2012, doi: 10.1007/s10909-012-0505-y
\bibitem{2016SPIE.9914E..0VH}
    Design of 280 GHz feedhorn-coupled TES arrays for the balloon-borne polarimeter SPIDER, J.~Hubmayr et al., 2016, doi: 10.1117/12.2231896
\bibitem{2016SPIE.9914E..16S}
    The design and characterization of wideband spline-profiled feedhorns for Advanced ACTPol, S.~Simon et al., 2016, doi: 10.1117/12.2233603

\bibitem{2008PhRvD..77h3003S}
    CMB polarization systematics due to beam asymmetry: Impact on inflationary science, M.~Shimonet al., 2008, doi:10.1103/PhysRevD.77.083003
\bibitem{2015Duff}
    Duff, S. , Hilton, G. , Hubmayr, J. , Beall, J. , Austermann, J. , Becker, D. and Van, J. (2015), Advanced ACTPol Multichroic Polarimeter Array Fabrication Process for 150 mm Wafers, Journal of Low Temperature Physics, [online], https://tsapps.nist.gov/publication/get\_pdf.cfm?pub\_id=919372
\bibitem{2016Hubmayr}
    Design of 280 GHz feedhorn-coupled TES arrays for the balloon-borne polarimeter SPIDER, J.~Hubmayr et al., 2016, doi: 10.1117/12.2231896

\bibitem{2016Shibo}
    Development of octave-band planar ortho-mode transducer with kinetic inductance detector for LiteBIRD, S.~Shibo et al., 2016, doi: 10.1117/12.2231877

\bibitem{2010Niemack}
    ACTPol: a polarization-sensitive receiver for the Atacama Cosmology Telescope, M.D.~Neamack et al., 2010, doi :10.1117/12.857464
\bibitem{2021Abitbol}
    The Simons Observatory: gain, bandpass and polarization-angle calibration requirements for B-mode searches, Abitbol et al., 2021, doi: 10.1088/1475-7516/2021/05/032
\bibitem{2018Yates}
    Eliminating stray radiation inside large area imaging arrays, Yates et al. 2018, doi: 10.1117/12.2315045

\end{thebibliography}
\end{document}